\documentclass[final,5p,times,twocolumn]{elsarticle}

\biboptions{authoryear}

\usepackage{mathtools, % instead amsmath, provide `\mathclap`
            nccmath, % for smaller `sum` symbol 
            amssymb}  

\usepackage{xspace} % handles spaces after newcommands
\usepackage{float} % use this to enable [H] at a figure to place it exactly HERE
\usepackage[separate-uncertainty=true]{siunitx} % display all units in the same way
\usepackage{textcomp} % enables the uses of \textit{}
\usepackage{hyperref} % hyperlinks to URLs, use \href{url}{abbreviation} or \url{url}
\hypersetup{colorlinks=true}
\usepackage{caption} % needed for usage of caption*{}
\usepackage{subcaption} % needed for subfigures
\usepackage{booktabs} % for toprule, midrule, bottomrule in tables
\usepackage{multirow} % for multi rows in tables
\usepackage{rotating} % to use sidewaystable which rotates the table
\usepackage{pdflscape}

\usepackage{graphicx} % handles pathes to images, \graphicspath, includegraphics,...
\graphicspath{{../general_documentation/figures}} % path points to general documentation

\usepackage{hyperref}

\newcommand{\integral}{\textit{INTEGRAL}\xspace}
\newcommand{\xmm}{\textit{XMM-Newton}\xspace}
\newcommand{\xmmns}{\textit{XMM-Newton}}
\newcommand{\chandra}{\textit{Chandra}\xspace}
\newcommand{\chandrans}{\textit{Chandra}}

\newcommand{\einsteinns}{\textit{Einstein}}
\newcommand{\exosatns}{\textit{EXOSAT}}

\newcommand{\arielvns}{\textit{Ariel V}}
\newcommand{\swift}{\textit{Swift}\xspace}
\newcommand{\swiftns}{\textit{Swift}}

\newcommand{\rosat}{\textit{ROSAT}\xspace}
\newcommand{\rosatns}{\textit{ROSAT}}

\newcommand{\heaoI}{\textit{HEAO-1}\xspace}
\newcommand{\uhuru}{\textit{Uhuru}\xspace}
\newcommand{\uhuruns}{\textit{Uhuru}}
\newcommand{\velaVb}{\textit{Vela 5B}\xspace}
\newcommand{\velaVbns}{\textit{Vela 5B}}

\newcommand{\nustarns}{\textit{NuSTAR}}
\newcommand{\erosita}{\textit{eROSITA}\xspace}
\newcommand{\nicer}{\textit{NICER}\xspace}
\newcommand{\maxi}{\textit{MAXI}\xspace}
\newcommand{\bepposax}{\textit{BeppoSAX}\xspace}
\newcommand{\rxte}{\textit{RXTE}\xspace}
\newcommand*\aap{A\&A}

\newcommand*\aj{AJ}

\newcommand*\apj{ApJ}

\newcommand*\apjs{ApJS}

\newcommand*\mnras{MNRAS}

\newcommand*\nat{Nature}

\newcommand*\prl{Phys.~Rev.~Lett.}

\newcommand{\fluxunits}{{erg s$^{-1}$cm$^{-2}$ }}

\newcommand{\papertwo}{Paper II\xspace}

\journal{Astronomy and Computing}

\begin{document}

\begin{frontmatter}




\title{HILIGT, Upper Limit Servers I - Overview\tnoteref{label1}}
\tnotetext[label1]{\url{http://xmmuls.esac.esa.int/hiligt}}


\author[1]{Saxton, R.D.}
\ead{richard.saxton@sciops.esa.int}
\author[2]{K{\"o}nig, O.}
\author[3]{Descalzo, M.}
\author[4]{Belanger, G.}
\author[4]{Kretschmar, P.}
\author[4,5]{Gabriel, C.}
\author[6]{Evans, P.A.}
\author[7]{Ibarra, A.}
\author[1]{Colomo, E.}
\author[8]{Sarmiento, M.}
\author[7]{Salgado, J.}
\author[7]{Agrafojo, A.}
\author[9]{Kuulkers, E.}

\address[1]{Telespazio U.K. Ltd. for the European Space Agency (ESA), European Space Astronomy Centre (ESAC), Camino Bajo del Castillo s/n, 28692 Villanueva de la Ca{\~{n}}ada, Madrid, Spain}
\address[2]{Dr.~Karl-Remeis-Sternwarte and Erlangen Centre for Astroparticle Physics, Friedrich-Alexander-Universit\"at Erlangen-N\"urnberg, Sternwartstr. 7, 96049 Bamberg, Germany}
\address[3]{Department of Electrical Engineering, Technical University of Denmark (DTU), Ørsteds Plads 326, 2800 Kongens Lyngby, Denmark}
\address[4]{European Space Agency (ESA), European Space Astronomy Centre (ESAC), Camino Bajo del Castillo s/n, 28692 Villanueva de la Ca{\~{n}}ada, Madrid, Spain}
\address[5]{Committee on Space Research (COSPAR) – 2, place Maurice Quentin, 75039 Paris Cedex 01, France}
\address[6]{University of Leicester, X-ray and Observational Astronomy Group, School of Physics and Astronomy, University Road, Leicester, LE17RH, UK}
\address[7]{QUASAR Science Resources for the European Space Agency (ESA), European Space Astronomy Centre (ESAC), Camino Bajo del Castillo s/n, 28692 Villanueva de la Ca{\~{n}}ada, Madrid, Spain}
\address[8]{Aurora Technology B.V for the European Space Agency (ESA), European Space Astronomy Centre (ESAC), Camino Bajo del Castillo s/n, 28692 Villanueva de la Ca{\~{n}}ada, Madrid, Spain}
\address[9]{European Space Agency (ESA), European Space Research and Technology
Centre (ESTEC), Keplerlaan 1, 2201 AZ Noordwijk, The Netherlands}

\begin{abstract}
The advent of all-sky facilities, such as the Neil Gehrels Swift observatory, the All Sky Automated Search for Supernovae (ASAS-SN), \erosita and Gaia has led to a new appreciation of the importance of transient sources in solving outstanding astrophysical questions. Identification and catalogue cross-matching of transients has been eased over the last two decades by the Virtual Observatory but we still lack a client capable of providing a seamless, self-consistent, analysis of all observations made of a particular object by current and historical facilities. HILIGT is a web-based interface which polls individual servers written for \xmm, \integral and other missions, to find the fluxes, or upper limits, from all observations made of a given target. These measurements are displayed as a table or a time series plot, which may be downloaded in a variety of formats. HILIGT currently works with data from X-ray and Gamma-ray observatories.

%The \ult allow to generate long-term lightcurves of historical X-ray data and
%calculate upper limits at any celestial position via a web
%interface. They utilize data from 12 satellites including current
%observatories such as \xmm and \integral, back to \rosat, \exosat,
%\einstein and even \arielv and \uhuru. This paper aims on describing the
%detailed software layout, technical aspects as well as each individual
%mission. We describe our choices on catalogue calls,
%footprint calculations and database handling.
\end{abstract}

\begin{keyword}
Catalogs \sep Surveys \sep X-rays: general \sep Instrumentation: detectors \sep Upper limit \sep Aperture photometry



\end{keyword}

\end{frontmatter}



% Introduction of general_documentation gives a broad introduction to
% X-ray astronomy. This is not necessary for this paper. Introduction
% should give a short overview of the software, essentially the
% summary of paper I by Richard
\section{Introduction}
Since the first discovery of X-ray emission from SCO~X-1 in 1962 by the Aerobee 150 sounding rocket \citep{Giaconni62}, X-ray telescopes have been flown on satellites on a regular basis. Their observations provide a record of the flux emitted from the brightest sources which now stretches back for 50 years. Detected sources from the majority of these missions are provided in specific catalogues \citep[e.g.][]{Evans10,Webb20,Evans_swift} or are served by the comprehensive  multi-mission archive hosted at HEASARC \citep{Heasarc}\footnote{\url{https://heasarc.gsfc.nasa.gov/}}.
Many missions have also left a legacy of sky images. These allow upper limits to fluxes to be found from sky positions where a source has not been detected and catalogued.
%Much of the X-ray sky is variable on different timescales. 

As new sources are detected, due to the
introduction of more sensitive instruments or because they have transitioned into
a brighter state, it is of interest to know how previous
observations of that position on the sky can constrain the history of the flux. For known
sources, while targeted observations are often covered
by catalogues, serendipitous non-detections and detections of lower flux may not be. To fully exploit the datasets of all X-ray missions, 
functionality is needed capable of returning catalogue values or of calculating upper limits from images. In this way, historical light curves can be generated using the full dataset of each mission.  

\begin{figure}
  \includegraphics[width=1.0\textwidth]{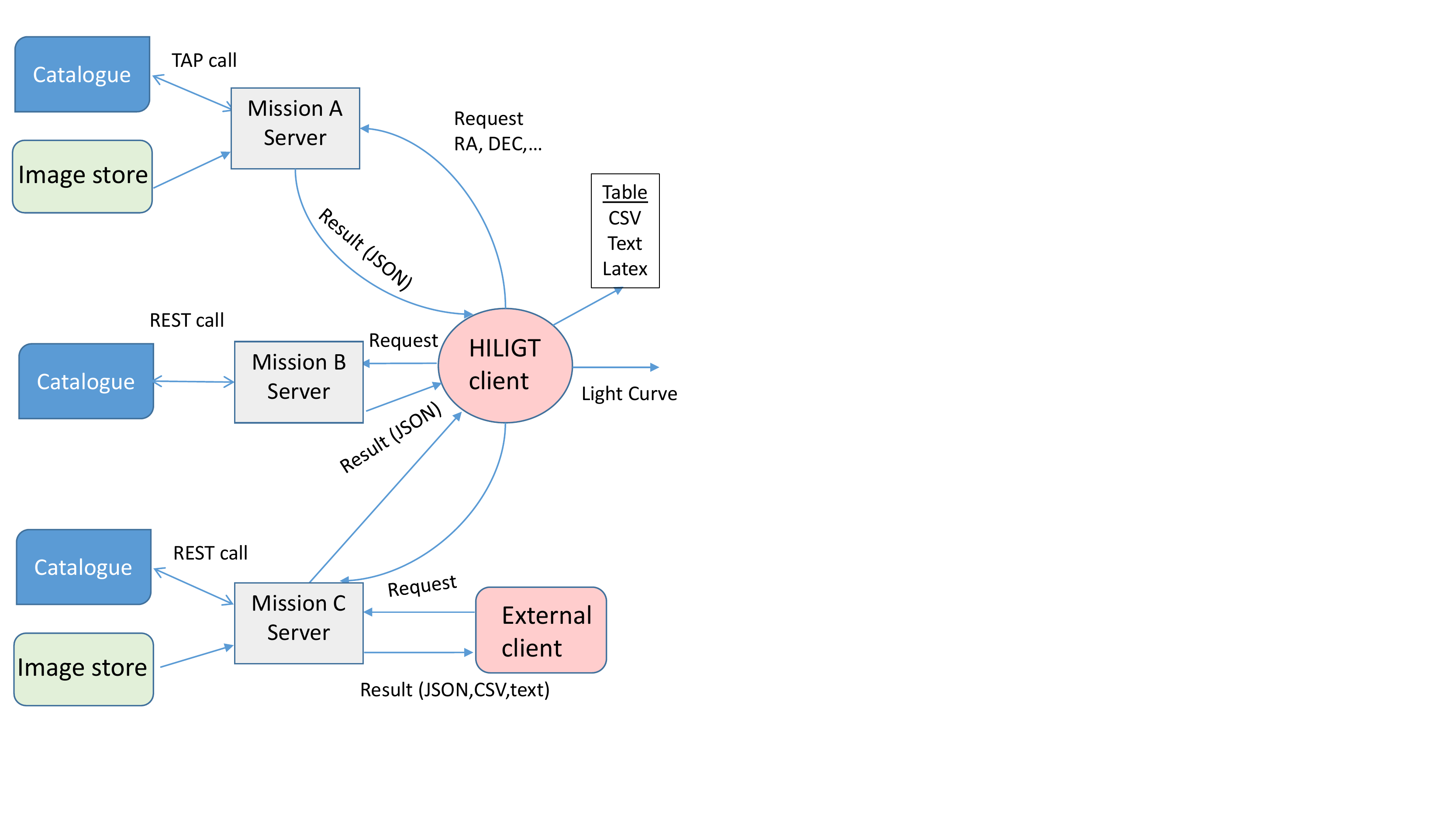}
\caption{Information flow between servers and clients.}
\label{fig:servers_clients}       
\end{figure}

To this end a suite of flux / upper-limit servers have been written, one for each mission, which run in
parallel and either forward on results from the best available on-line catalogue or analyse the images and return upper limits from a given position (Fig.~\ref{fig:servers_clients}). 
%This allows the full dataset of each mission to be exploited and in particular facilitates rapid detection of transients in new data.

In this paper we review the design and describe the input parameters and the fields returned by each server. A companion paper (K\"{o}nig et al. 2021; hereafter \papertwo) describes in detail how each individual server has been constructed. A brief summary of the missions which are currently supported is given in Table~\ref{tab:missions}. Results are returned as a Representational State Transfer (REST) service which may be called by any suitable client; a web client with plotting facilities has been produced and is described in detail in Sect.~\ref{sec:clients}. The collection of servers and the web client together are called the HIgh-energy LIght curve GeneraTor (HILIGT) which replaces an earlier upper limit server hosted at the \xmm  science operations centre (SOC) that worked exclusively with \xmm data\footnote{A comprehensive upper limit server dedicated to the \xmm EPIC cameras called FLIX is available at \url{https://www.ledas.ac.uk/flix/flix.html}}.

\begin{table*}[ht]
{
\caption{Currently supported missions}
\label{tab:missions}      % is used to refer this table in the text
\begin{center}
\begin{small}
\begin{tabular}{llccll}
\hline\hline                 % inserts double horizontal lines
% \multicolumn{7}{c}{emission model} & \multicolumn{3}{c}{intrinsic absorption} & C/dof  \\
\\
%\multicolumn{2}{c}{Power-law$^{a}$}  & Distant Reflection$^{b}$ & \multicolumn{3}{c}{zxipcf~$^{c}$} & \multicolumn{2}{c}{Ionized Reflection$^{d}$} & C/dof~$^{e}$ \\
%$\Gamma$ & Norm  &  Norm & $N_{H}$ & $xi$ & cf & $xi$ & Norm & \\
 %        & keV$^{-1}$ cm$^{-2}$ s$^{-1}$ & keV$^{-1}$ cm$^{-2}$ s$^{-1}$ & 10$^{22}$ cm$^{-2}$ & log & \% & log & cm$^{-2}$ s$^{-1}$ & \\
Mission & Instrument(s) & modes$^{a}$ & Data source$^{b}$ & Period & Energy range$^{c}$  \\
        &               &       &    & (year) & (keV) \\
\hline\noalign{\smallskip}
VELA 5B & ASM   & survey & CAT & 1969-1979 & 3--12 \\
Uhuru   &       & survey & CAT & 1970-1973 & 2--20 \\
Ariel-V & ASM, SSI & survey & CAT & 1974-1980 & 2--18 \\
Heao-1  &  A1   & pointed & CAT & 1977-1979 & 0.25--25 \\
Einstein & IPC, HRI & pointed           &     CAT, UL     &  1978-1981 & 0.15--4.0 \\
EXOSAT  & LE, ME           & slew, pointed     &  CAT, UL & 1983-1986 & 0.05--50.0 \\
Ginga   &  LAC & pointed & CAT & 1987-1991 & 1.5--30 \\
ROSAT    &    PSPC, HRI     &  survey, pointed & CAT, UL & 1990-1998 & 0.1--2.4 \\
ASCA    & GIS, SIS & pointed & CAT & 1993-2000 & 0.4--12 \\
INTEGRAL & ISGRI &  pointed & UL & 2000-2021+  &  20--100 \\
\xmm  &  EPIC-pn  &  slew, pointed & CAT, UL & 2000-2021+  &  0.2--12 \\
Swift  & XRT  &  pointed  & UL & 2005-2021+ & 0.2--10.0 \\
\noalign{\smallskip}\hline
\end{tabular}
\\
\end{small}
\end{center}
$^{a}$ Missions operate in survey, slew and/or pointed observation mode.\\
$^{b}$ Data source can be CATalogue and/or calculated Upper Limit (UL).\\
$^{c}$ Full energy range of the instrumentation.\\
}
\end{table*}

\section{Mission count rate and flux servers}
The fundamental property measured by a photon counting detector is the background-subtracted count rate. This can be converted into an energy flux by convolving with the efficiency and spectral resolution of the detector as a function of energy and by assuming a spectral model. %i.e. a flux $F_{b}$ in the energy band $E_{0} - E_{1}$, is related to the count rate by:

%\begin{equation}
%    F_{b} = C_{r} \int_{E_{0}}^{E_{1}} A_{E}M_{E}
%\end{equation}

%where $A_{E}$ is the effective area of the detector system in cm$^{2}$ as a function of energy, E, and $M_{E}$ is the luminosity contained in the spectral model in units of ergs s$^{-1}$. 

Each HILIGT server either returns the source count rate and error with the equivalent flux derived from the input spectral model, or calculates the upper limit on the source count rate, at a user-selectable 1,2 or 3-$\sigma$ level, with the equivalent upper limit on the flux. 

To enable fluxes to be compared between missions a standard set of energy bands have been defined: 0.2--2 keV (soft), 2--12 keV (hard) and 0.2--12 keV (total). The one current exception is \integral which uses the three bands 20--40, 40--60 and 60--100 keV. These ranges have been set as a compromise
based on the actual energy bands of the mission instrumentation (see Table~\ref{tab:missions}).

\begin{figure}
  \includegraphics[width=1.0\textwidth]{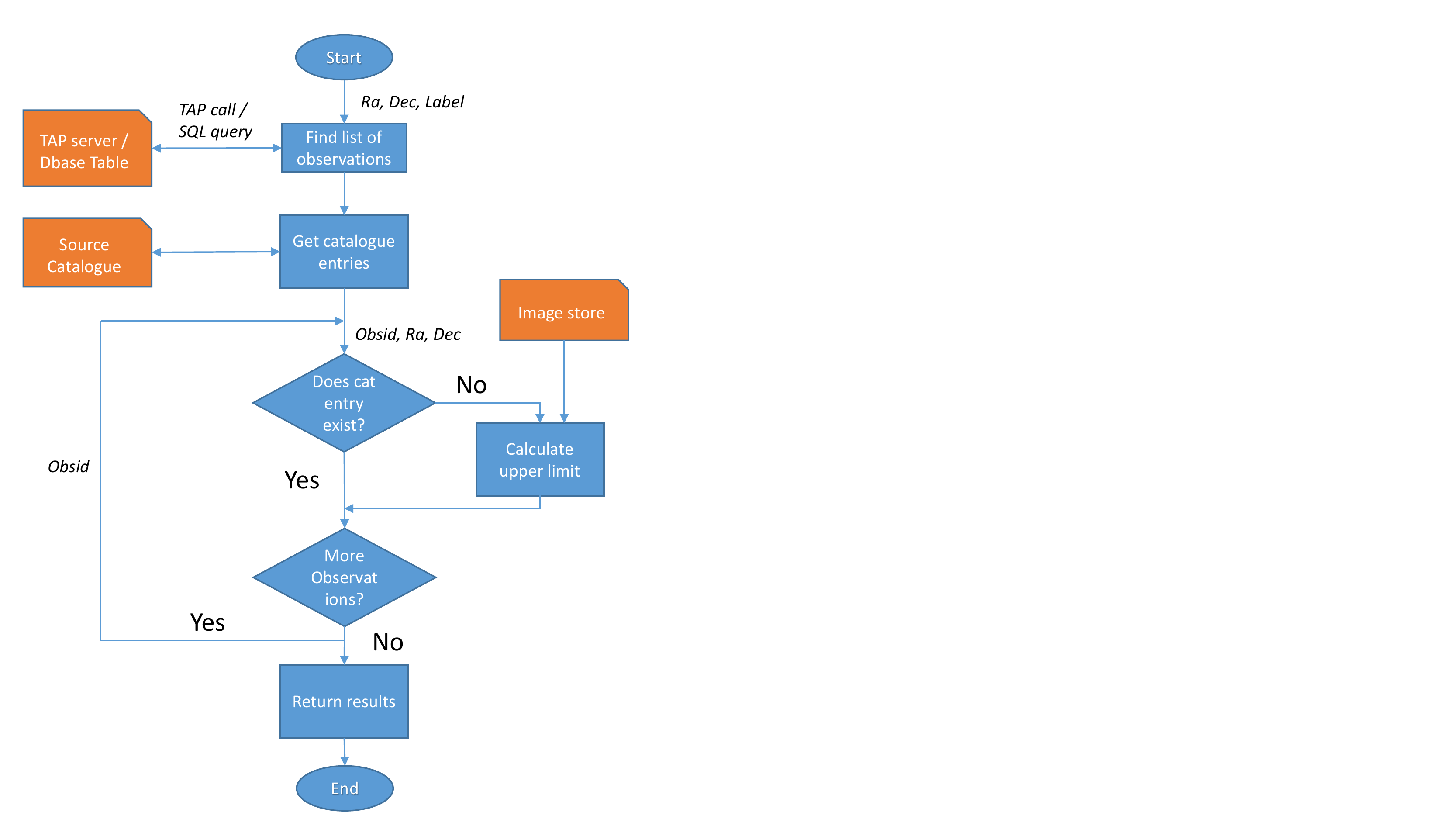}
\caption{Flowchart of the design of a mission flux / upper limit server.}
\label{fig:flowchart}       
\end{figure}

The processing flow of a server is given in Fig.~\ref{fig:flowchart} and consists of the following steps.

\begin{itemize}
\item{Find all observations which contain a given set of equatorial coordinates (RA, Dec). Ultimately this is done by searching a database for image footprints which contain the sky position.
The derivation of the footprints for each mission is described in \papertwo and the comparison is made using the pgsphere package\footnote{\url{https://pgsphere.github.io}}, which is in some cases, for example for the \xmm slew and pointed data, provided by a Table Access Protocol \cite[TAP;][]{Dowler2019}} call.

\item{Get catalogue entries for this position. Catalogue fields, principally the source count rate and error, are accessed by making a TAP call \footnote{\url{http://nxsa.esac.esa.int/tap-server/tap}} for \xmm pointed data or by using the w3Browse facility provided at HEASARC
\footnote{\url{https://heasarc.gsfc.nasa.gov/W3Browse/w3browse-help.html}. An alternative approach would have been to use the
HEASARC TAP service.}. Cross-matching with catalogues is carried out using a radius which takes into account the systematic positional error of the given catalogue and the likelihood of source confusion. Specific details are given in \papertwo, Section 3.10.
}

\item{Calculate an upper limit for the sky position in images where a catalogue entry is not available, using
aperture photometry.
%The calculation is performed with the {\em eupper} task within the XMM science analysis system \citep[SAS; ][]{Gabriel06}, using images, and if available background images and exposure maps, which are stored on a local disk. 
%The calculation finds source counts from a circle about the sky position and background counts from an annulus about the circle, with radii which depend on the point spread function (PSF) of the observatory and its pointing accuracy. An arbitrary decision has been taken to return a flux if the background-subtracted count-rate is $\geg2$ times the error and otherwise return an upper limit.

This calculation relies on knowledge of the instrument vignetting and the Point Spread Function (PSF) which is used to calculate the encircled energy function (EEF; the fraction of counts which lie within 
the extraction region). In order to provide a practical distinction between a detection and a non-detection, a decision has been taken to return a flux if the background-subtracted count rate is $\geq2$ times the error and otherwise return an upper limit at a user-selectable confidence level of 1, 2 or 3 sigma. The upper limit calculation uses Bayesian statistics based on the algorithm given in \cite{Kraft:1991a}. We give a
description of this calculation in \papertwo. 
}
\item{Calculate the flux from the count rate. This is done by multiplying by a factor, based on a provided spectral model. While in practise the spectral model of
an astrophysical object can be very complicated, for simplicity in passing parameters, the model is currently limited to either a
 power-law with index, $\Gamma=0.5,1.0,1.5,1.7,2.0,2.5,3.0,3.5$, or a black-body of temperature $kT=60,100,300,1000$ eV, attenuated by a neutral absorber of column, $N_{H}=1\times10^{20}, 3\times10^{20}, 1\times10^{21}$ cm$^{-2}$. These values sample the typical observed spectra of sources and include the spectrum
 used to calculate fluxes in the \xmm catalogue ($\Gamma=1.7$, $N_{H}=3\times10^{20}$ cm$^{-2}$, which is based on the average X-ray spectrum of an AGN \citep[e.g.][]{turnerpounds89}).
 Conversion factors have been hard-coded into the server software for efficiency and were originally calculated using the PIMMS package \citep{Mukai93} or inferred from the relevant literature 
 \citep{Forman1978,Kaluzienski1977,Warwick1981,Wood1984,Nugent1983}}.

\end{itemize}

An alternative approach would be to pre-calculate and store fluxes for each catalogued source, for each spectral model, and make them available to HILIGT through a TAP server.
While this would result in a performance improvement it would prove rather inflexible, as updates would be needed if the default spectral models were changed or if our understanding of the instrument calibration changed. It would also be unviable should the system be updated to allow the user to choose a spectral model on-the-fly.

\subsection{Inter-mission cross-calibration}
Significant community effort is put into the cross-calibration of X-ray detectors by the International Astronomical Consortium for High-Energy Calibration \citep[IACHEC; ][]{Sembay10}. 
Recent missions, such as \xmmns, \swiftns, \nustarns, SUZAKU and \chandra are found to return consistent fluxes to within 10--15\% \citep{Madsen17}. For the older missions, the factors to convert count rate into flux tend to be based on observations of the Crab nebula (see \papertwo), which has a power-law spectrum of $\Gamma=2.1$. Fluxes are then only strictly valid for similar spectral shapes and furthermore the Crab flux and spectral shape are known to vary at the level of a few percent \citep{Madsen2017}. To compensate for this, in the returned flux for these missions a systematic error 
of magnitude 30\% (\velaVbns), 15\% (\arielvns), 20\% (\heaoI) and 20\% (\uhuruns) is added in quadrature to the statistical error.

As technology has improved, there has been a tendency for the sensitivity of missions to increase and the beam size to decrease. 
This leads to quite different levels of output between the missions, e.g. \velaVb has data for 99 sources, while the \xmm point source catalogue contains almost 900,000 sources. A summary of the sky coverage and number of catalogued sources is given in \papertwo (Table 1).

\subsection{Input / output parameters}

A definition of the input parameters accepted by the server is given in Table~\ref{tab:inputpar}.
As a minimum, a server needs to receive the coordinates of a sky position. A default spectral model of a power-law of spectral slope, $\Gamma=2$, absorbed by a column of neutral hydrogen of $3\times10^{20}$ cm$^{-2}$ is used for flux conversion and two-sigma upper limits are returned by default.

Note that for a meaningful comparison of flux between different missions it is essential that the
spectral model used for the flux conversion is an accurate representation of the spectrum of
the source in question. The {\em wrong} spectral model can artificially introduce strong apparent
variability between missions \citep[e.g.][]{Page15}.

The server can be made to ignore catalogue values and recalculate results by setting the parameter {\em usecat=NO}. 

Each mission server returns a Javascript Object Notation (JSON) structure containing the full set of fields 
described in Table~\ref{tab:outputpar}, or comma-separated values (CSV) or text records containing a reduced, easy to display, output structure. 
The output fields contain the count rate, 
number of source counts, number of background counts, exposure time, flux calculated using the supplied model and ancillary information.

%To facilitate the comparison of flux between missions, three fiducial energy bands: soft (0.2--2 keV),
%hard (2--12 keV) and total (0.2--12 keV) have been defined\footnote{Except INTEGRAL which uses the bands (20--40, 40--60 and 60--100 keV).}. 
Count rates returned by the
instrumentation, which as can be seen from Table\ref{tab:missions} may be recorded over differing 
energy ranges (e.g. 0.3--2 keV and 2--10 keV for the Swift-XRT telescope) are converted into fluxes in
the standard bands. Upper limits are available for all missions which
have produced a database of images that can be analysed. Upper limits have been pre-calculated for the INTEGRAL Soft Gamma-ray Imager (ISGRI) 
and stored in a database table (see \papertwo).

\subsection{Clients}
\label{sec:clients}
Servers may be hosted anywhere, although they are currently all
located at the European Space Astronomy Centre (ESAC)\footnote{Note that the Swift-XRT server forwards
the request onto a further server hosted within the Leicester Database \& Archive Service (LEDAS) which actually makes the calculation.} and for convenience have
been made accessible via a single REST call:

%\begin{verbatim}
   \url{http://xmmuls.esac.esa.int/ULSservice_passthru?MISSION=<mission>&ra=<ra>&dec=<dec>}
%   \end{verbatim}
See Table~\ref{tab:inputpar} for the full set of input parameters.
Any server compliant with the input and output parameters detailed in Tables~\ref{tab:inputpar} and ~\ref{tab:outputpar}
could be added.

\begin{figure*}
  \includegraphics[width=1.0\textwidth]{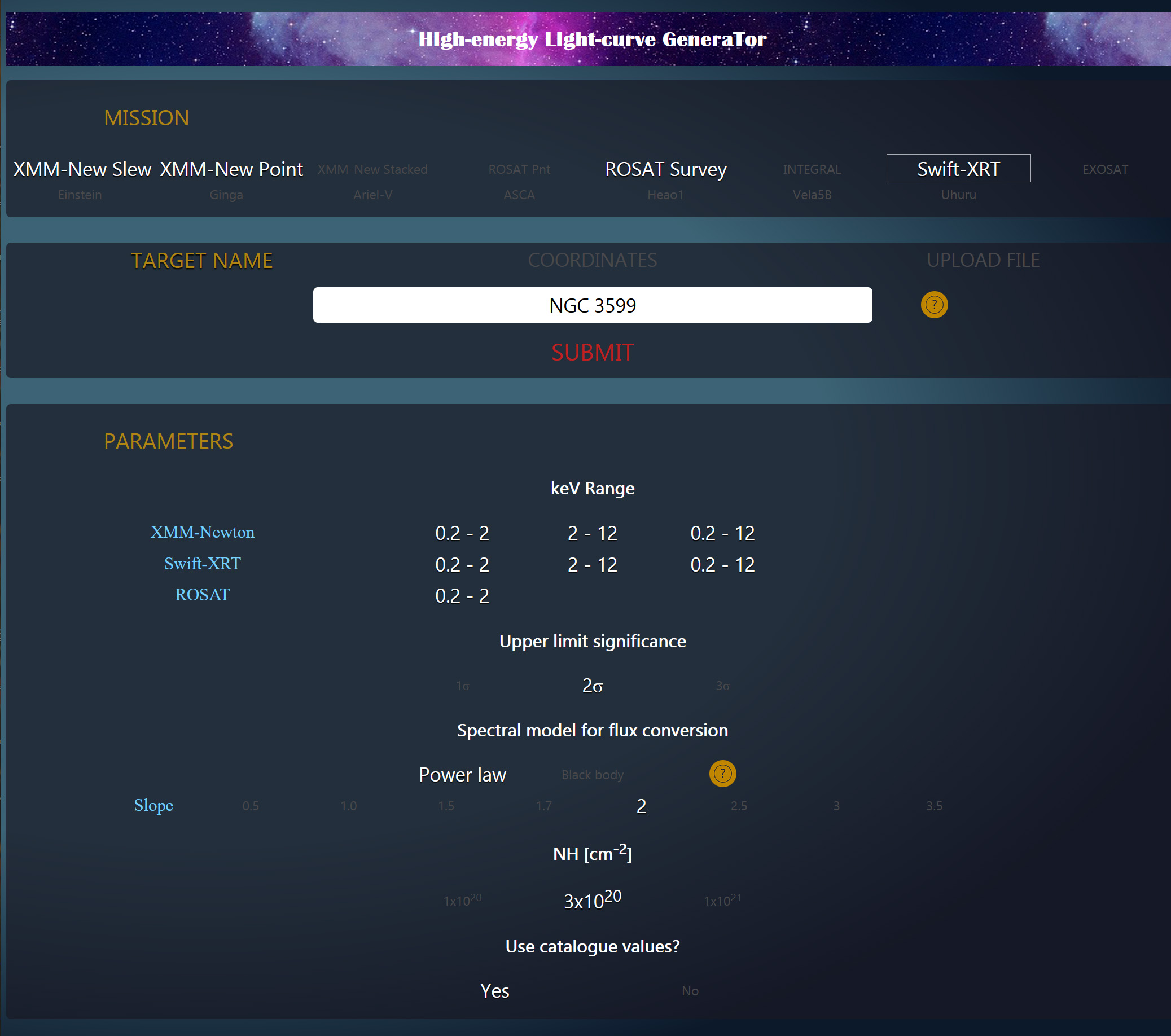}
\caption{Input page for the HILIGT web client. Clicking on a mission selects it for processing, turning it from grey to bold white.}
\label{fig:web_input}       
\end{figure*}

An example terminal client written in Python is available for download from:

%\begin{verbatim}
   \url{http://xmmuls.esac.esa.int/hiligt/scripts/hiligt.py}
%   \end{verbatim} 
   
In addition an interactive, web-based client written in Javascript is available at:

%\begin{verbatim}
   \url{http://xmmuls.esac.esa.int/hiligt}
%\end{verbatim}

The entry page of the web client (Fig.~\ref{fig:web_input}) allows sky positions to be entered as a coordinate pair, a target name, to be resolved by SIMBAD, or a text file containing a list of coordinates.
The missions to be interrogated can be selected from the upper panel (Fig.~\ref{fig:web_input}). Standard energy bands, the spectral model for conversion of count rate to flux, the confidence level of the upper limit and whether to use catalogue
values, may all be selected. 

\begin{figure*}
  \includegraphics[width=1.0\textwidth]{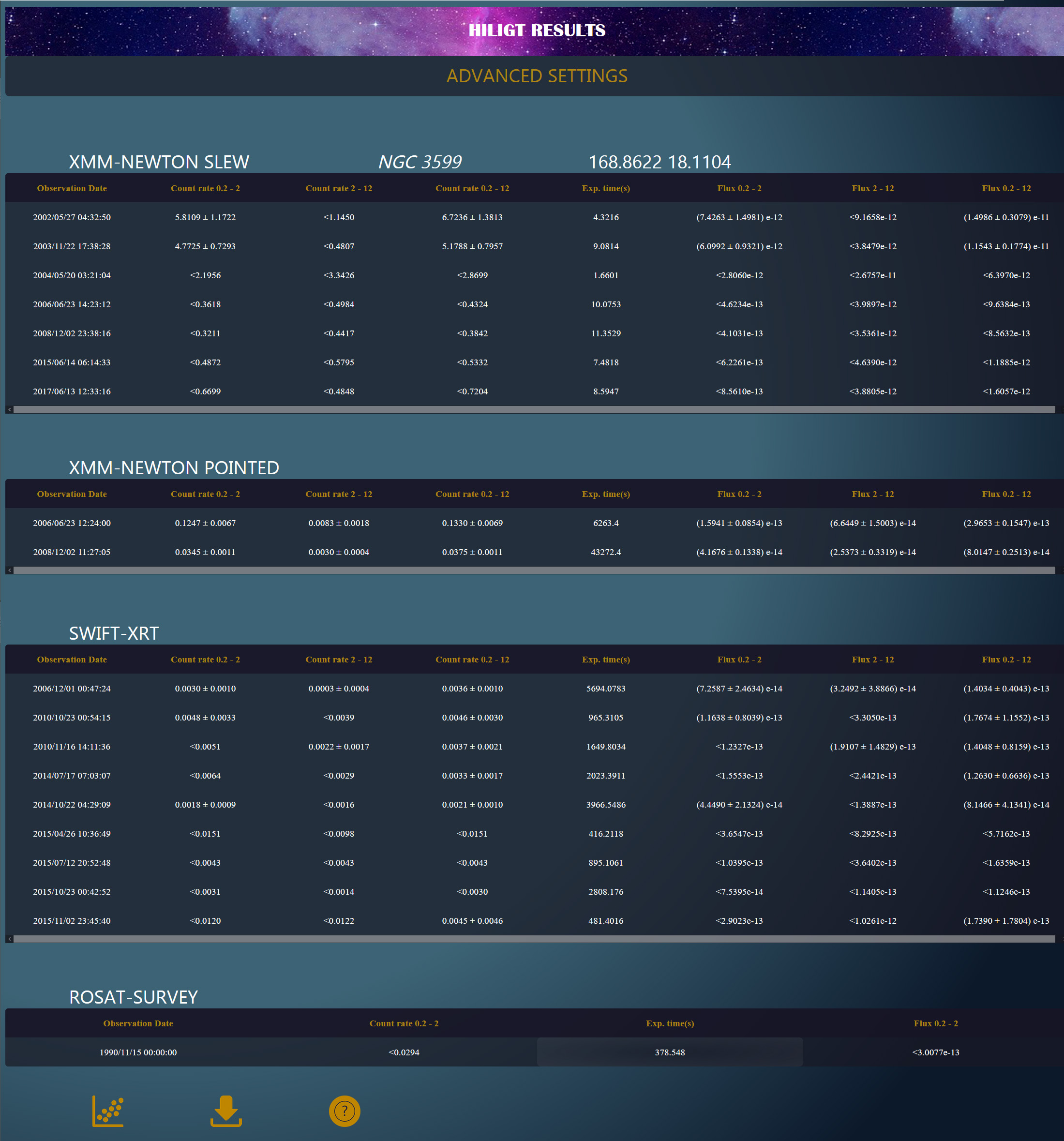}
\caption{Output page for the HILIGT web client.}
\label{fig:web_output}       
\end{figure*}

The web client returns a basic output page of results for each source position and mission, listing the observation date, count rate and fluxes for each 
observation and selected energy-band (Fig.~\ref{fig:web_output}). By clicking on "Advanced Settings" further fields may be selected for display from a top panel. From the output page the results may be saved into a text file, a csv file
or downloaded into a  \LaTeX \xspace table ready to be inserted into a paper. 
A basic light curve plotting function, based on the
ZingChart v2.8.6 tool\footnote{\url{http://www.zingchart.com}}, may be invoked from the output page. Alternatively, a Python code to plot the results with greater flexibility may be used.\footnote{
\url{http://xmmuls.esac.esa.int/hiligt/scripts/lightcurve.py}
}

\section{Scientific Uses}

HILIGT lends itself to long timescale variability analysis, to the search
for transients and to broad-band spectral variability analysis. We briefly
describe an example of each below.

\subsection{Analysis of long-term light curves}

\begin{figure*}
  \includegraphics[width=1.0\textwidth]{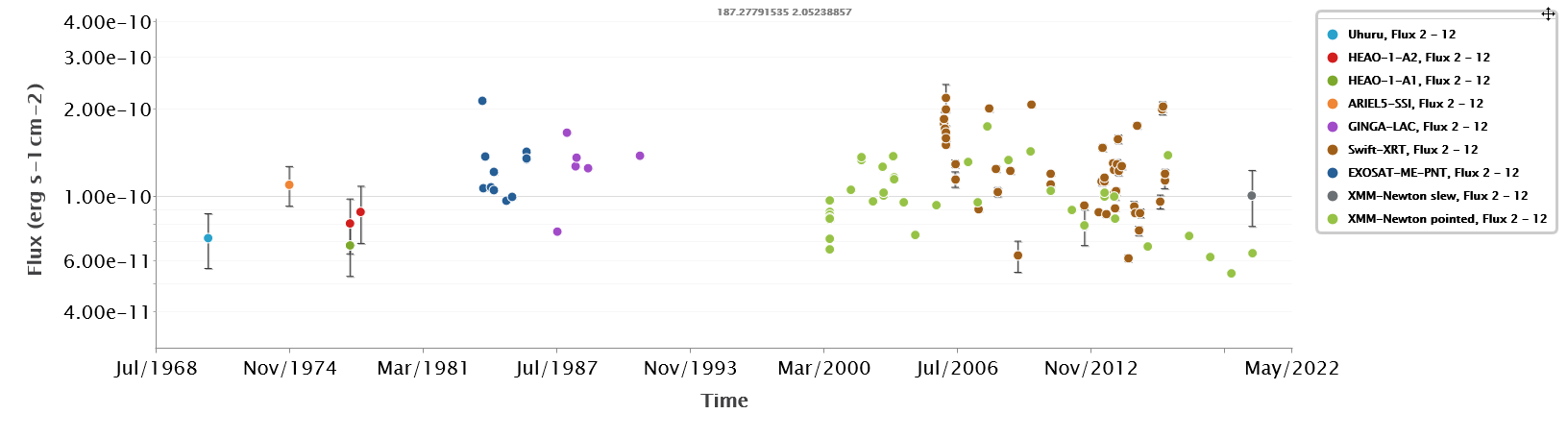}
\caption{Light curve of the 2--12 keV flux from the QSO 3C~273 using the
\uhuru, Heao-1, Ariel-V, EXOSAT-ME, Ginga, Swift and \xmm flux servers.
Flux conversion has been performed with a model of an absorbed power-law
of slope 1.7 and galactic column of $N_{H}=1\times10^{20}$ cm$^{-2}$.
The plot has been produced by the
ZingChart graphics tool integrated into HILIGT.}
\label{fig:lc_3c273}       
\end{figure*}

The first quasar to be identified was 3C~273 in 1963 \citep{Schmidt63}. This was first 
observed in X-rays by the satellite \uhuru in 1970 and has subsequently been a scientific or calibration target of all the major X-ray observatories, resulting in a rich dataset which covers its activity over the last 50 years. 
In Fig.~\ref{fig:lc_3c273} we show the time series of the 2--12 keV flux from  3C~273 observed between 1970 and 2020, extending
 the work of \cite{Soldi08} who concentrated on observations taken between 1985 and 2006. 
The light curve is quite stable, having maintained a mean 2--12 keV  
flux of $1.0\times10^{-10}$ \fluxunits over the 50 year span (45 years in the source
restframe; z=0.158) with a peak-to-peak variation of a factor 4.
%between $5.5\times10^{-11}$ and $2.2\times10^{-10}$\fluxunits.  
We note that the minimum flux 
of $5.5\times10^{-11}$\fluxunits was recorded on
2019-07-02 being $\sim20$\% fainter than the previous minimum registered in July 2015 \citep{Kalita17}. 
%This stability is most likely a consequence of the large mass of the black hole ($M_{BH}\sim6\times10^{9}$ \msolarns) in the nucleus of this galaxy \citep[e.g.][]{Kalita15}. 
This long time series gives confidence that fluxes from old and new missions can sensibly be compared despite the widely differing beam sizes of the instrumentation.

\subsection{Transient search}

\begin{figure*}
  \includegraphics[width=1.0\textwidth]{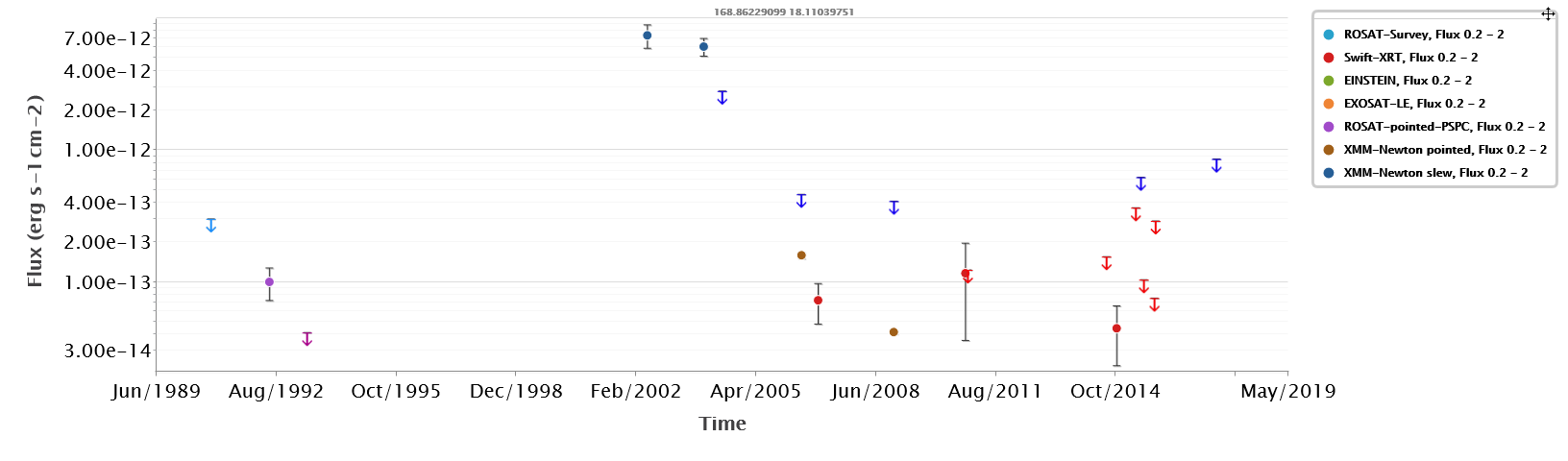}
\caption{Full light curve of the 0.2--2 keV flux from the candidate tidal disruption event NGC~3599 using the
\rosat, \swift and \xmm flux servers.
Count rates have been converted into flux using a spectral model of a kT=60 eV black-body absorbed by a galactic column of $N_{H}=1\times10^{20}$ cm$^{-2}$ 
\citep[see][]{Esquej:2008a,Saxton:2015a}.
%Both plots have been produced by the ZingChart graphics tool integrated into HILIGT
}.
\label{fig:lc_3599}       
\end{figure*}

To know whether a new X-ray source is an interesting transient it is necessary to compare the measured flux with previous detections or upper limits. It is straightforward to write a script to extract and compare historical measurements at a given sky position from a set of the HILIGT flux and upper limit servers. An example plot for the light curve of the galaxy NGC~3599 is shown in Fig.~\ref{fig:lc_3599}. This candidate tidal disruption event (TDE) \citep{Esquej:2008a,Saxton:2015a} exhibited an increase in flux of greater than a factor 100
between a \rosat observation of 1993 and an \xmm slew observation made in 2002,
before decaying back to a quiescent state.

\subsection{Search for spectral variability}

\begin{figure*}
  \includegraphics[width=1.0\textwidth]{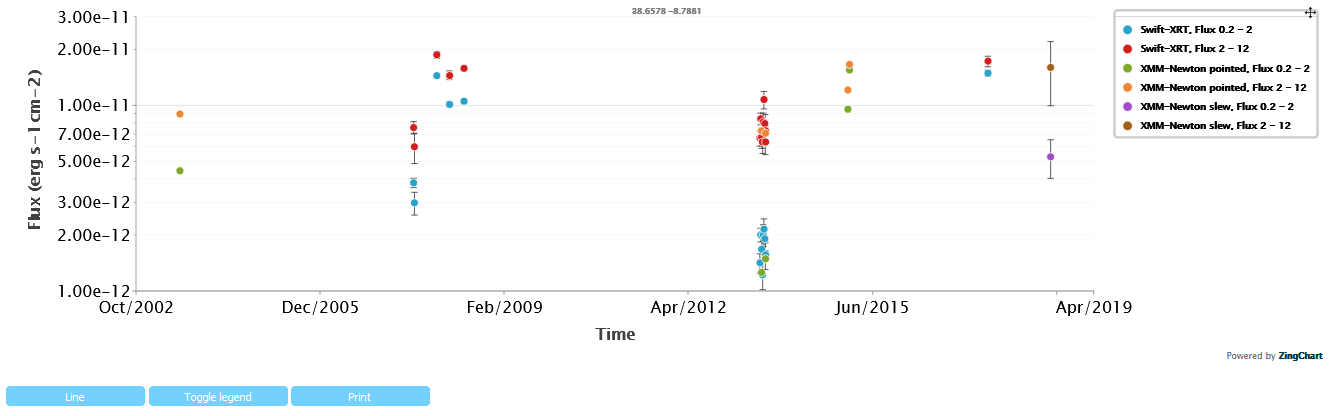}
\caption{Light curve of the 0.2--2 keV (green, blue and purple points) and 2--12 keV (orange, red and brown points) flux from the AGN NGC 985 using the \xmm and \swift flux servers.} 
%The plot has been produced by the ZingChart graphics tool integrated into HILIGT

\label{fig:lc_ngc985}       
\end{figure*}

By plotting the soft and hard energy band long-term light curves a simple check for 
spectral variability can be achieved. In Fig.~\ref{fig:lc_ngc985} we show the 
light curve of the Seyfert I galaxy, NGC~985 in the 0.2--2 and 2--12 keV bands. It can be seen that the
source spectrum hardened considerably in 2013, due to the soft flux diminishing while the hard flux remained stable at historical levels. This spectral change has been attributed to the development of a multi-component warm absorber around 2013 which preferentially absorbed the soft X-rays \citep{Ebrero16}. 

\section{Access}

While the extraction of count rates from catalogues is quite quick, the calculation of upper limits from images
is a slower process. This limits the size of queries which can practically be issued to the system and, for the moment, the web client only accepts input files with up to ten sky positions. Larger queries may be built by using clients which poll the individual servers one source at a time.

A new strategy to increase the efficiency of the system by pre-calculating and storing \xmm upper limits \citep[RapidXMM;][]{Ruiz2021} has been developed and will be integrated into HILIGT in the near future.
This will greatly increase the throughput of the system and allow large queries to be processed for the \xmm servers.

%Also places to download the software and/or links

% Should contain a conclusion, summary and short outlook what needs to
% be done in the future for Release II (right now: soft link)
\section{Conclusion and future plans}

We have presented a system (HILIGT) for finding the flux recorded at a particular sky position
for all X-ray observations. The system consists of a set of RESTful servers which may be 
called by any suitable client. A web-based client, capable of saving the historical 
flux in a number of formats and plotting the results has been introduced.
The servers and client were made public on September 9th 2019.

The science available from HILIGT would benefit greatly from the addition of flux servers for current missions such as \chandrans, \nustarns, \nicer, \maxi and \erosita as well as
extension to earlier missions such as \rxte and
\bepposax. 

The system for pre-calculating and storing \xmm upper limits should be extended to other commonly used
missions (\rosatns, \exosatns, \einsteinns) to maximise the efficiency and usefulness of the system.

Finally, we note that HILIGT may prove useful for cross-calibration studies between X-ray missions.

\section{Acknowledgments}
We acknowledge support from ESA through the Faculty of the European Space Astronomy Centre (ESAC) - Funding reference 560/2019 and under the ESAC trainee program of 2018.

\clearpage
%\bibliographystyle{elsarticle-num} 
%\bibliography{references.bib}

%\printbibliography


%\begin{thebibliography}{00}
%\end{thebibliography}


\clearpage
%\appendix

%Below we describe the fields which comprise the input parameters and output fields of a mission server.

\begin{table*}[ht]
{\small
\caption{Input parameters of a server query.}
\label{tab:inputpar}      % is used to refer this table in the text
\begin{center}
\begin{small}
\begin{tabular}{lccccl}
\hline\hline                 % 
Field & Data type & Units & range & Default & Description \\
 &  &  & or options & & \\
\hline\noalign{\smallskip}
mission & string & - & - & - & Name of mission$^{a}$ \\
ra  &  double precision  &  degrees  &  0.0 - 360.0 & - & Right ascension\\
dec &  double precision  &  degrees  &  -90.0 - +90.0 & - & Declination\\
label & string           &     -     &  - & - & Source name \\
band  & string           &     -     &  soft,hard,total,all & all & Energy band(s) \\
model & string           &     -     &  plaw,bbody & plaw & spectral model \\
specparam & float  &  - / keV  & options$^{b}$ & 2.0 or 100eV & slope or temperature \\
nh   & float & cm$^{-2}$ & 1.0E20, 3.0E20, 1.0E21 & 3.0E20 & Column density \\
ulsig & int & sigma & 1,2,3 & 2 & Upper limit significance \\
usecat & bool & - & "yes","no" & "yes" & Use catalogue value? \\
FORMAT & string & - & "text","text/html","csv","JSON" & "text" & Output format \\
\noalign{\smallskip}\hline
\end{tabular}
\\
\end{small}
\end{center}
$^{a}$ Supported missions are: XMMpnt, XMMslew, XMMStacked, RosatSurvey, RosatPointedPSPC, RosatPointedHRI, Integral, ExosatLE, ExosatME, Einstein, Ginga, Ariel5, Heao1, SwiftXRT, Asca, Uhuru, Vela5B.\\ 
$^{b}$ Currently supported values are: slope=0.5,1.0,1.5,1.7,2.0,2.5,3.0,3.5 or black-body temperature=0.06,0.1,0.3,1.0 keV.
}
\end{table*}

%\begin{itemize}
%    \item     "ra": "221.52", 
%    \item      "dec": "68.9585", 
%    \item     "label": "",  
%    \item   "band": "all,soft,hard,total", 
%    \item "model": "plaw,bbody",
%    \item "specparam" : "plaw slope or black-body %temperature (keV)",
%    \item "nh" : "Absorption column cm$^{-2}$",
%    \item "usecat" : "true,false"
%\end{itemize}

%Below we describe the fields which comprise the output of a server record.

\begin{table*}[ht]
{\small
\caption{Output fields of a server record.}
\label{tab:outputpar}      % is used to refer this table in the text
\begin{center}
\begin{small}
\begin{tabular}{lcccl}
\hline\hline                 % inserts double horizontal lines
% \multicolumn{7}{c}{emission model} & \multicolumn{3}{c}{intrinsic absorption} & C/dof  \\
\\
%\multicolumn{2}{c}{Power-law$^{a}$}  & Distant Reflection$^{b}$ & \multicolumn{3}{c}{zxipcf~$^{c}$} & \multicolumn{2}{c}{Ionized Reflection$^{d}$} & C/dof~$^{e}$ \\
%$\Gamma$ & Norm  &  Norm & $N_{H}$ & $xi$ & cf & $xi$ & Norm & \\
 %        & keV$^{-1}$ cm$^{-2}$ s$^{-1}$ & keV$^{-1}$ cm$^{-2}$ s$^{-1}$ & 10$^{22}$ cm$^{-2}$ & log & \% & log & cm$^{-2}$ s$^{-1}$ & \\
Field & Data type & Units & range & Description \\
 &  &  & or options & \\
\hline\noalign{\smallskip}
start\_date &  string  &  -  &  - & Start of obs - e.g. 2004-11-19T04:34:09\\
end\_date &  string  &  -  &  - & End of obs - e.g. 2004-11-19T05:47:35\\
crate &  float  &  counts / second  &  - & Count rate\\
crate\_err &  float  &  counts / second  &  - &  Count rate error\\
crate\_flux &  float  &  \fluxunits  &  - & Source flux\\
crate\_flux\_err &  float  &  \fluxunits  &  - &  Error on source flux\\
ul &  float  &  counts / second  &  - &  Upper limit on count rate\\
ul\_flux &  float  &  counts / second  &  - & Upper limit of flux\\
ulsig &  int  &  -  &  - &  Sigma of upper limit\\
obsid &  string  &  -  &  - & Observation identifier\\
src\_counts &  int  &  counts  &  - &  Counts in source region\\
bck\_counts &  int  &  counts  &  - &  Counts in bckgnd region\\
exptime &  float  &  seconds  &  - &  Exposure time\\
filt &  string  &  -  &  - &  Filter\\
image &  string  &  -  &  - &  Name of image\\
bkgimage &  string  &  -  &  - &  Name of background image\\
expmap &  string  &  -  &  - &  Name of exposure map\\
eef &  float  &  -  &  - &  Encircled energy fraction\\
elow &  float  &  keV  &  - &  Lowest energy of band\\
ehigh &  float  &  keV  &  - &  Highest energy of band\\
mission &  string  &  -  &  - &  Name of mission, e.g. XMM-Newton\\
instrum &  string  &  -  &  - &  Name of instrument, e.g. EPIC-pn\\
ra  &  double precision  &  degrees  &  0.0 - 360.0 & Right ascension\\
dec &  double precision  &  degrees  &  -90.0 - +90.0 & Declination\\
label & string           &     -     &  - & Source name \\
model & structure$^{a}$  &     -     & - & spectral model \\
status & string & -     & "OK","NoData" & status for this instrument \\
\noalign{\smallskip}\hline
\end{tabular}
\\
\end{small}
\end{center}
$^{a}$ This is a JSON structure containing: Column density (cm$^{-2}$), spectral model
("plaw" or "bbody") and spectral parameter (power-law index or black-body temperature in keV).
}
\end{table*}

%\begin{itemize}
%    \item start\_date: "2004-11-19T04:34:09",
%    \item "end\_date": "2004-11-19T05:47:55",
%    \item "crate": null, 
%    \item "crate\_err": null, 
%    \item "crate\_flux": " ", 
%    \item "crate\_flux\_err": " ", 
%    \item   "ul": "0.5447", 
%    \item "ul\_flux":6.9287e-13, 
%    \item "ulsig": "2"
%    \item     "obsid": "9090600002", 
%    \item "src\_counts": "1",    
%    \item "bck\_counts": "0", 
%     \item "exptime" "10.569572", 
%    \item "filt": "Medium", 
%    \item "image":"P9090600002PNS003IMAGE\_6036.FTZ",  %  
%    \item "bkgimage": "NotSet", 
%    \item "expmap": "P9090600002PNS003EXPMAP6036.FTZ",
%    \item        "eef": "0.845", 
%    \item        "elow": 0.2,    
%    \item "ehigh": 2.0,   
%    \item "instrum": "EPIC-pn",
%    \item     "ra": "221.52", 
%    \item      "dec": "68.9585", 
%    \item     "label": "",  
%    \item   "mission": "XMM-Newton slew", 
%    \item "model": [
%        3e+20, 
%        "plaw", 
%        2.0
%    ], 
%   \item  "status": "Ok",
%\end{itemize}

\end{document}